\RequirePackage{fix-cm}
\documentclass[twocolumn]{svjour3}          
\smartqed  
\usepackage{graphicx}
\usepackage[colorinlistoftodos]{todonotes}
\usepackage{multirow}
\usepackage{hyperref}
\usepackage{array}
\usepackage{tabu}
\usepackage{tabularx}
\newcolumntype{M}[1]{>{\centering\arraybackslash}m{#1}}
\newcolumntype{R}[1]{>{\raggedleft\arraybackslash}m{#1}}
\newcolumntype{L}[1]{>{\arraybackslash}m{#1}}

\begin{document}
\sloppy

\title{Harvesting the Public MeSH Note field
}
\subtitle{Identifying the previous state of new descriptors in the MeSH thesaurus as Supplementary Concept Records
}


\author{Anastasios Nentidis$^{1,2}$ \and Anastasia Krithara$^1$ \and Grigorios Tsoumakas$^2$ \and Georgios Paliouras$^1$}


\authorrunning{Nentidis \textit{et al.}} 


\institute{
$^1$National Center for Scientific Research ``Demokritos'', Athens, Greece \\
\email{\{tasosnent, akrithara, paliourg\}@iit.demokritos.gr}\\
  $^2$Aristotle University of Thessaloniki, Thessaloniki, Greece\\
\email{\{nentidis, greg\}@csd.auth.gr}\\
}

\maketitle

\begin{abstract}
In this document, we report an analysis of the Public MeSH Note field of the new descriptors introduced in the MeSH thesaurus between 2006 and 2020. 
The aim of this analysis was to extract information about the previous status of these new descriptors as Supplementary Concept Records. 
The Public MeSH Note field contains information in semi-structured text, meant to be read by humans. 
Therefore, we adopted a semi-automated approach, based on regular expressions, to extract information from it. 
In the large majority of cases, we managed to minimize the required manual effort for extracting the previous state of a new descriptor as a Supplementary Concept Record. 
The source code for this analysis is openly available on GitHub.\footnote{\url{https://github.com/tasosnent/MeSH_Extension}}

\keywords{MeSH \and terminology extension \and semantic indexing \and biomedical literature
}
\end{abstract}

\section{Introduction}

The analysis reported here took place in the context of a broader study of the semantic provenance of new descriptors in the MeSH thesaurus~\cite{nentidis2021all}. 
In this study, the new descriptors were organized into specific provenance categories and types. In particular, the new descriptors were classified into one of four provenance categories based on the explicit or implicit coverage of their topic on MeSH prior to their introduction by predecessor descriptors, which are called Previous Hosts (PHs).  
In addition, the new descriptors were also grouped into six provenance types, based on their current hierarchical relation to their PHs. 

The Public MeSH Note (PMN\footnote{\url{http://id.nlm.nih.gov/mesh/vocab\#publicMeSHNote}}) field is a field of the descriptor records in MeSH. It consists of sentences that are usually separated by semicolons and may provide varying information such as the year of introduction of the descriptor and other relevant descriptors. 
Of particular interest for this work are some PMN sentences reporting that a previous Supplementary Concept Record (SCR) \textit{X}, corresponding to this descriptor, \textit{was indexed under} some descriptors \textit{Y} in specific periods. This is useful as in some cases an SCR promoted to descriptor may undergo some minor term modifications and receive a new identifier. In such cases, the PMN is the only source of information for identifying the old SCR for the new descriptor, which would otherwise be considered totally new. The fact that \textit{X} in such cases has been an SCR was first communicated to us by researchers in NLM\footnote{Cho, Dan-Sung (NIH/NLM) personal communication}. Later, this rule was also confirmed by the results of the analysis.

\section{The analysis}

\begin{figure*}    
\center
\includegraphics[width=0.81\textwidth]{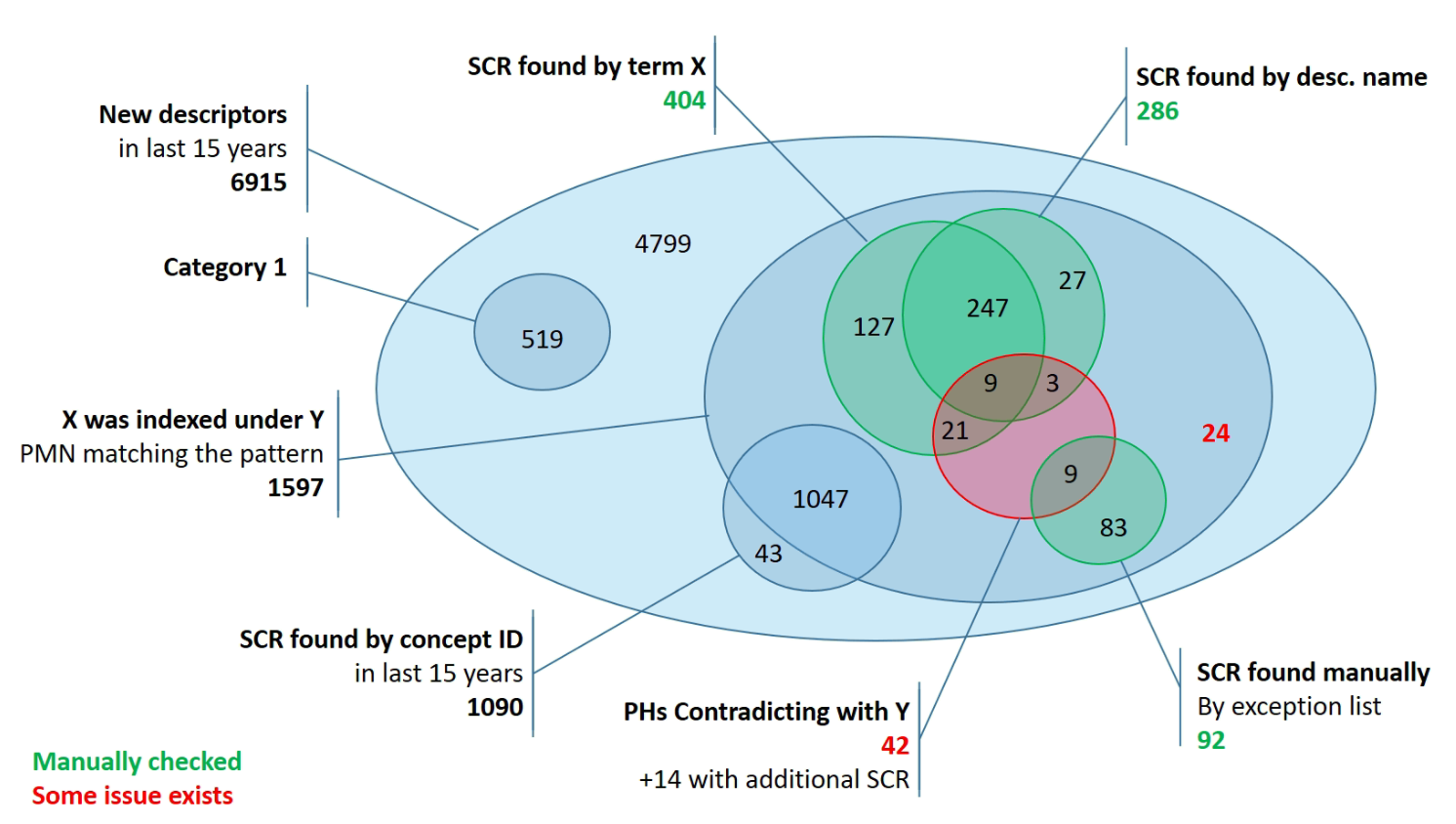}
\caption{Harvesting the Public MeSH Note field of new MeSH descriptors introduced in the last fifteen years. Each circle represents a set of descriptors based on whether and how the PMN field was considered for identifying the previous status of the descriptor as a Supplementary Concept Record. }
\label{fig:pmn}       
\end{figure*}

Figure~\ref{fig:pmn} provides an overview of this PMN analysis of the new descriptors introduced in the last fifteen years, which is also summarized in Table~\ref{tab:sum}. 
Out of the 6915 new descriptors identified for the last 15 years, 519 belong to Category 1 as their preferred concept was previously available as a subordinate concept. The PMN field was examined for the remaining 6396 of them in order to identify whether they had a previous status as SCRs (Category 2) or not (Category 3 or Category 4). In particular, it turned out that 1597 of them had a PMN filed matching the \textit{X was indexed under Y} pattern. At this point, we examined only whether the PMN field contains the sub-string ``was indexed under'', regardless of the consistency of the structure of \textit{X} and \textit{Y} elements. This was done in order to have a better estimation of how many cases should be added in Category 2, not affected by the successful extraction of the terms in \textit{X} or \textit{Y} and their mapping into SCRs or descriptors.

\begin{table*}
\caption{Sets of new descriptors introduced in the last fifteen years in the context of harvesting their Public MeSH Note field.}
\label{tab:sum}       
\centering
\begin{tabular}{ll}
\textbf{Descriptor set}   & \textbf{Number of descriptors}  		\\
\hline
All new descriptors   & 6915 		\\
Category 1    & 519  		\\
non Category 1    & 6396		\\
PMN not covered by the ``...was indexed under...'' pattern  & 4799		\\
PMN covered by the ``...was indexed under...'' pattern & 1597		\\
SCR found overall & 1616		\\
SCR found by pref Concept & 1090 \\
SCR found by pref Concept covered by the pattern & 1047 \\
PMN covered by the pattern and not found by concept & 550 \\
SCR found by term & 404 \\
SCR found by descriptor Name & 286 \\
SCR found by both term and name & 256 \\
SCR found by exception & 92 \\
SCR not found  & 24 \\
SCR not found by other means than perf Concept & 526 \\
\end{tabular}
\end{table*}

For the majority (1047, ~66\%) of these 1597 cases of new descriptors amenable to the pattern, we were able to identify the previous SCR based on direct concept identifier comparison. In particular, about 96\% of the cases found by identifier comparison were also amenable to the pattern, which is an additional confirmation that the pattern can be considered as a general rule. 
For the remaining 550 cases that were amenable to the pattern but not resolved by concept identifier comparison, an approach, based on regular expressions, was developed for automatically extracting from the PMN the preferred term \textit{X} of the old SCR. This worked directly for 404 cases, identifying the term of the old SCR and mapping it to an actual SCR object of the corresponding MeSH version. 

Examining the remaining 146 cases, we noticed that\textit{ X} can be empty in some cases. For example, Calbindin 2 (D064032) has a PMN filed with value ``2014; was indexed under CALCIUM-BINDING PROTEINS, VITAMIN D-DEPENDENT 1987-2013''. For exploiting such cases, we used the current descriptor name as the preferred term to map in an SCR of the corresponding MeSH version. This approach identified the old SCR for 30 additional cases. In addition, this approach also worked for most (256, ~63\%) of the 404 cases already resolved by the PMN-extracted X term, leading, for all these cases, to the selection of the same SCR, which reveals the consistency of these two alternative approaches.

For the remaining 116 cases where no SCR was found to match exactly either to the \textit{X} term or the descriptor name, the automated method employed two alternative approaches for finding non-exact matches to suggest potential old SCRs for manual selection as described below:
\begin{itemize}
    \item \textit{Partial matching}:
     The first approach splits the query terms (i.e. extracted X and/or descriptor name) and the potential matching SCR terms into parts by any non-alphanumeric character. Then, it looks for SCR terms containing all the parts of the query term(s) with the same number of occurrences. These SCR terms are ideal candidates for matching SCRs. For example, the concept “adrenomedullin receptor” (C093200) is a good candidate SCR for the descriptor “Receptors, Adrenomedullin” (D058265).
    In the absence of such a candidate SCR consisting of the same parts as the query term(s), candidate SCRs that include all the parts of the query term(s) are considered, even if they contain some additional parts. For example, the concept ``ATXN3 protein, human'' (C092341) is a good candidate SCR for the descriptor ``Ataxin-3'' (D000067699) because it includes the term ``ataxin-3 protein, human'' that matches the parts ``Ataxin'' and ``3'' with the additional parts ``protein'' and ``human''.
   \item \textit{String similarity}:
   The second approach is based on the Levenstein distance of the query term(s) to the potential matching SCR terms, suggesting the ones with the smallest distance. For example, the SCR ``cryptochrome'' (C063074) is a candidate SCR for the descriptor ``Cryptochromes'' (D056931).
\end{itemize}

All the candidates for both the \textit{X} term or the descriptor name, when available, were inspected manually. When possible, the best candidate was selected, adding the old SCR for 92 more cases and leaving only 24 unresolved cases without mapping to an SCR.

Finally, the previous hosts (PH) identified for all the 537 cases of old SCRs extracted from the PMN note, either directly with exact match or by examining suggested alternatives, were manually inspected in comparison to the corresponding PMN (actually the \textit{Y} part) to estimate their agreement. This cross-validation procedure revealed 42 cases with non-identical PHs identified compared to the ones reported in the PMN note:
\begin{itemize}
    \item 12 cases with some different PHs.
    \item 12 cases with all the PHs reported in PMN and some additional.
    \item 18 cases with PHs reported in PMN only, but not all of them.
\end{itemize}

As the PMN field has been observed to contain various inconsistencies, such as typos (e.g. ``2006; CATACHOLAMINE TRANSPORT PROTEIN was indexed...'' instead of ``catecholamine'' in D050482) and cases of missing elements (e.g. ``2012; HLA-DRB5 was indexed under 1992-2011'' in D059847), we keep the PHs extracted from the identified SCR as the most reliable. 

\section{Results}

In conclusion, the procedure left only 24 non-resolved cases, which corresponds to less than 5\% of the 550 cases on which it has been applied. These 24 cases form a loss of less than 2\% of the total number of the 1616 Category 2 descriptors. In addition, as already discussed in PH cross-validation experiment above, the PMN field often includes inconsistencies and typos. Therefore, these 24 non-resolved cases may indeed correspond to rare cases of SCRs that are in effect beyond the scope of the current conceptual framework. For example, the descriptor ``Interleukin-4 Receptor alpha Subunit'' (D053662) has a PMN ``2007; CD124 ANTIGENS was indexed under RECEPTORS, INTERLEUKIN-4 1998-2005'' which reports that it an SCR was used for this entity two years prior to its introduction (until 2005). As a result, in the last previous version of MeSH, in 2006, neither the ``CD124 ANTIGENS'' SCR nor the ``RECEPTORS, INTERLEUKIN-4'' descriptor were available and the meaning was actually not explicitly covered as such. Therefore, it is reasonable to annotate it as a new concept.

\begin{acknowledgements}
This research work was supported by the Hellenic Foundation for Research and Innovation (HFRI) under the HFRI Ph.D. Fellowship grant (Fellowship Number: 697).
We are grateful to James Mork and Dan-Sung Cho from the National Library of Medicine (NLM) for kindly providing valuable feedback on this work.
\end{acknowledgements}

%
%

\bibliographystyle{spmpsci}      

\bibliography{main}



\end{document}